\documentclass[desactivate]{aa}  

\usepackage{natbib}
\bibpunct{(}{)}{;}{a}{}{,}

\usepackage{graphicx}
\usepackage{revsymb}	
\usepackage{amsmath}
\usepackage{txfonts}
\usepackage{amssymb}
\usepackage{geometry} 
\usepackage[parfill]{parskip} 
\usepackage{slantsc}
\usepackage{array}
\usepackage{amsfonts}
\usepackage{epstopdf}	
\usepackage{epsfig}	
\usepackage{url}
\usepackage[colorlinks=true,linkcolor=black, urlcolor=black, citecolor=black]{hyperref}
\usepackage{color}
\usepackage{xfrac}

\usepackage{sidecap}
\usepackage[font=small, labelfont=bf]{caption}
\usepackage{subcaption}

\usepackage{multirow}
\usepackage{booktabs}
\usepackage{threeparttable}

\begin{document}
   \title{Influence of the magnetic activity cycle on mean density and acoustic radius inversions}
\author{J. B\'{e}trisey\inst{1} \and D.~R. Reese\inst{2} \and S.~N. Breton\inst{3} \and A. -M. Broomhall\inst{4} \and A.~M. Amarsi\inst{1} \and R.~A. Garc{\'\i}a\inst{5} \and O. Kochukhov\inst{1}}
\institute{Department of Physics and Astronomy, Uppsala University, Box 516, SE-751 20 Uppsala, Sweden\\email: 	\texttt{jerome.betrisey@physics.uu.se}
\and LIRA, Observatoire de Paris, Université PSL, Sorbonne Université, Université Paris Cité, CY Cergy Paris Université, CNRS, 92190 Meudon, France
\and  INAF – Osservatorio Astrofisico di Catania, Via S. Sofia 78, 95123 Catania, Italy
\and Centre for Fusion, Space and Astrophysics, Department of Physics, University of Warwick, Coventry CV4 7AL, UK
\and Université Paris-Saclay, Université Paris Cité, CEA, CNRS, AIM, 91191 Gif-sur-Yvette, France}
\date{\today}

\abstract
{Asteroseismic modelling is set to play a crucial role in upcoming space-based missions such as PLATO, CubeSpec, and Roman. Despite the significant progress made in this field, asteroseismology has uncovered notable discrepancies between observations and theoretical predictions. These discrepancies introduce non-negligible biases in stellar characterisation at the precision levels required by PLATO. Present modelling strategies typically disregard magnetic activity, assuming its impacts are concealed within the parametrisation of the so-called `surface effects'. However, this assumption has recently been challenged, as a significant imprint of magnetic activity on the asteroseismic characterisation of the Sun using forward modelling methods has been demonstrated.}
{Based on GOLF and BiSON observations of two full activity cycles of the Sun, a reference target for assessing the PLATO mission requirements, we quantified the impact of magnetic activity on solar mean density and acoustic radius inversions.}
{The GOLF and BiSON observations were segmented into yearly overlapping snapshots, each offset by 91.25 days. For each snapshot, we performed inversions to determine the mean density and acoustic radius. This approach enabled us to track the apparent temporal evolution of these two quantities and to estimate the systematic uncertainty associated with magnetic activity.}
{Similar to the findings obtained using forward methods, we observe a discernible imprint of the magnetic activity cycle on the solar mean density and acoustic radius as determined through helioseismic inversions. This imprint is consistent across both GOLF and BiSON datasets, and constitutes the largest source of systematic uncertainty in the solar asteroseismic characterisation. Additionally, the effects of magnetic activity are mitigated by the inclusion of low radial-order modes in the dataset, consistently with the literature, but we observe a significantly larger mitigation factor than previous measurements for other stellar variables such as the stellar age.}
{We recommend asteroseismic values for the solar mean density and acoustic radius: $\bar{\rho}_{\mathrm{inv}} = 1.4104\pm 0.0051$ g/cm$^3$ and $\tau_{\mathrm{inv}} = 3722.0\pm 4.1$ s. The suggested values correspond to the average over the two full activity cycles and the suggested uncertainties take into account the major sources of systematic errors, including the choice of physical ingredients in stellar models, stellar activity, and the surface effect prescription. We achieved a high precision of 0.36\% for the mean density and 0.11\% for the acoustic radius. These results are promising, as they demonstrate the potential to attain high precision levels for these quantities in Sun-like stars. A better-constrained mean density can be used to enhance the precision of the stellar radius, which is crucial for characterising exoplanetary systems. A more accurately determined stellar radius indeed leads to better estimates of the orbital distance and planetary radius of exoplanets.}


\maketitle

\defcitealias{Betrisey2023_AMS_surf}{JB23}

\section{Introduction}
A wide range of stellar oscillations are generated by the convective motions in the upper layers of solar-type stars. By analysing these oscillations, asteroseismology provides a unique window into the internal structure of stars, enabling us to determine their fundamental parameters -- such as mass, radius, age, and mean density -- with unparalleled precision and accuracy. Accurate stellar models are essential for comprehending planetary system evolution and tracing the history of our galaxy through Galactic archaeology \citep[see e.g.][for reviews]{Chaplin&Miglio2013,Garcia&Ballot2019,Aerts2021}. Following the success of missions like CoRoT \citep[Convection, Rotation and planetary Transits;][]{Baglin2009}, \textit{Kepler} \citep{Borucki2010}, K2 \citep{Howell2014}, and TESS \citep[Transiting Exoplanet Survey Satellite;][]{Ricker2015}, asteroseismic modelling is poised to play a pivotal role in upcoming space-based missions such as PLATO \citep[PLAnetary Transits and Oscillations of stars;][]{Rauer2024}, CubeSpec \citep{Bowman2022}, and Roman \citep{Huber2023}.

Despite remarkable achievements, asteroseismology has also revealed significant discrepancies between observed data and theoretical stellar models, leading to biases in stellar characterisation. This issue is becoming particularly important nowadays given the stringent precision requirements of the PLATO mission, which targets 15\% accuracy in mass, 1-2\% in radius, and 10\% in age for Sun-like stars (defined for a reference star with $\sim 6000$ K, $\sim 1M_\odot$, and $\sim 1R_\odot$). It should be noted that accurate stellar characterisation is essential not only for understanding the stars themselves but also for the precise characterisation of exoplanetary systems. This precision will enable accurate age dating of these systems, offering invaluable insights into their formation and evolution. Asteroseismic modelling faces three primary challenges: the physical ingredients used in stellar models \citep[e.g.][]{Buldgen2019f,Farnir2020,Betrisey2022}, surface effects \citep[e.g.][]{Ball&Gizon2017,Nsamba2018,
Jorgensen2020,Jorgensen2021,Cunha2021,Betrisey2023_AMS_surf}, and stellar magnetic activity \citep[e.g.][]{Broomhall2011,
Santos2018,PerezHernandez2019,Santos2019_sig,Santos2019_rot,Howe2020,Thomas2021,Santos2021,Santos2023,
Betrisey2024_MA_Sun}. The choice of physical ingredients in stellar models presents substantial challenges. For instance, the literature does not agree on the abundances used in the Sun \citep[see e.g.][and references therein]{Buldgen2023a,
Lind&Amarsi2024}, various opacity tables exist in the literature, and different formalisms can be used to describe the equation of state and microscopic diffusion. Surface effects stem from the limitations of 1D stellar evolutionary models in accurately handling convection in the near-surface layers where oscillations are excited, and the overlooking of non-adiabatic effects in most oscillation codes. As a consequence, frequency shifts, that increase with frequency, are detected between theoretical predictions and observations \citep[see e.g.][]{Kjeldsen2008}. Moreover, magnetic stellar activity can alter observed frequencies \citep[see e.g.][and references therein]{Garcia2024}, further complicating the situation. The current understanding of magnetic activity is incomplete, and distinguishing its effects from surface effects remains challenging. In the literature, frequency shifts due to magnetic activity are theorised to be caused by magnetic field variations in the sub-surface layers \citep[e.g.][]{Howe2002, Baldner2009} or a structural variation of the acoustic cavity \citep[e.g.][]{Woodard&Noyes1985,Fossat1987,Libbrecht&Woodard1990,Kuhn1998,Dziembowski&Goode2005,Basu2012}. Recent studies have shown that magnetic activity can significantly impact the estimation of stellar parameters such as mass, radius, age, and helium abundance in main-sequence stars \citep{Creevey2011,PerezHernandez2019,Thomas2021,Betrisey2024_MA_Sun}.

\citet{Betrisey2024_MA_Sun} explored the influence of magnetic activity on the asteroseismic characterisation of the Sun through forward modelling, revealing a noticeable effect of the solar activity cycle on the estimated seismic age. Building on this foundation, our study examines the impact of magnetic activity on the asteroseismic characterisation of the Sun using inverse techniques. Similar to \citet{Betrisey2024_MA_Sun}, our analysis utilises Doppler velocity observations from the Birmingham Solar Oscillations Network \citep[BiSON;][]{Davies2014,Hale2016} and the Global Oscillations at Low Frequencies \citep[GOLF;][]{Gabriel1995} instrument, encompassing solar cycles 23 and 24. In Sect. \ref{sec:datasets_and_modelling_strategy}, we introduce the datasets and elaborate on the modelling strategy for asteroseismic characterisation. Section \ref{sec:influence_magnetic_activity_cycle} focuses on assessing the impact of magnetic activity on mean density and acoustic radius inversions. Section \ref{sec:discussion} provides a quantitative analysis of various sources of systematic uncertainties and offers recommendations for uncertainty values to be adopted for the mean density and acoustic radius of the Sun as determined through seismic inversions. Finally, Sect. \ref{sec:conclusions} presents the conclusions of our study.

\section{Modelling strategy}
\label{sec:datasets_and_modelling_strategy}
In this study, we examine the influence of magnetic activity on the mean density and acoustic radius, as derived from seismic inversions. It should be noted that seismic data can be utilised through two principal methodologies \citep[see e.g.][for a recent review]{Buldgen2022c}. The first, called forward modelling, entails the computation of an evolutionary model along with its associated eigenfrequencies. This model is defined by several free parameters, including mass, age, initial hydrogen and helium fractions, and the overshooting parameter. These parameters are constrained by comparing the theoretical frequencies of the model with observational data. Typically, residual frequency differences persist between the forward model and the observations. The second uses seismic inversions that seek to exploit these residuals to apply minor structural corrections to the forward model. These inverse techniques are advantageous due to their data-driven nature, rendering them largely independent of the initial model assumptions. Furthermore, they were successfully applied to various asteroseismic targets \citep[see e.g.][]{DiMauro2004_theo,Buldgen2016b,Buldgen2016c,Buldgen2017c,
Bellinger2017,Buldgen2019b,Bellinger2019d,Buldgen2019f,Kosovichev&Kitiashvili2020,Salmon2021,Bellinger2021,Betrisey2022,
Buldgen2022b,Betrisey2023_rot,Betrisey2023_AMS_surf,Betrisey2024_AMS_quality,Betrisey2024_MA_Sun}.

\subsection{Datasets}
\label{sec:datasets}
We re-utilised the observational solar frequency series from GOLF and BiSON, as derived by \citet{Betrisey2024_MA_Sun}. The GOLF frequencies are derived from calibrated data following the procedure of \citet{Garcia2005} and extracted using the \textsc{apollinaire} software \citep{Breton2022_astro,Breton2022_helio}. The BiSON frequencies were extracted by following the methodology of \citet{Fletcher2009}. The series comprise 94 and 92 yearly overlapping snapshots, respectively, each delayed by a quarter of a year, fully covering solar cycles 23 and 24. Following \citet{Betrisey2024_MA_Sun}, we considered two primary mode sets: one including low radial-order modes (set~1) and one excluding them (set~2). Mode set~2 is typically expected for a G-type Sun-like star observed by PLATO, whereas a mode set similar to set~1 might be conceivable for the very best targets. PLATO will conduct photometric observations, which have a different background profile compared to the spectroscopic observations used in this study, complicating the detection of low-order oscillations. This is similar to the differences observed between modern spectroscopic GOLF data and photometric VIRGO/SPM data \citep[Variability of solar IRradiance and Gravity Oscillations / Sun PhotoMeters;][]{Frohlich1995}. Given that magnetic activity predominantly affects the highest frequency modes, it is worthwhile to test whether including lower radial-order modes could mitigate the influence of magnetic activity, akin to the observations of \citet{Betrisey2024_MA_Sun} using forward modelling. As an aside, we note that because the low radial-order modes are less heavily damped in our datasets, and therefore their peaks narrower, they tend to have the most precise frequencies as well.

\begin{figure}[t!]
\centering
\includegraphics[scale=0.55]{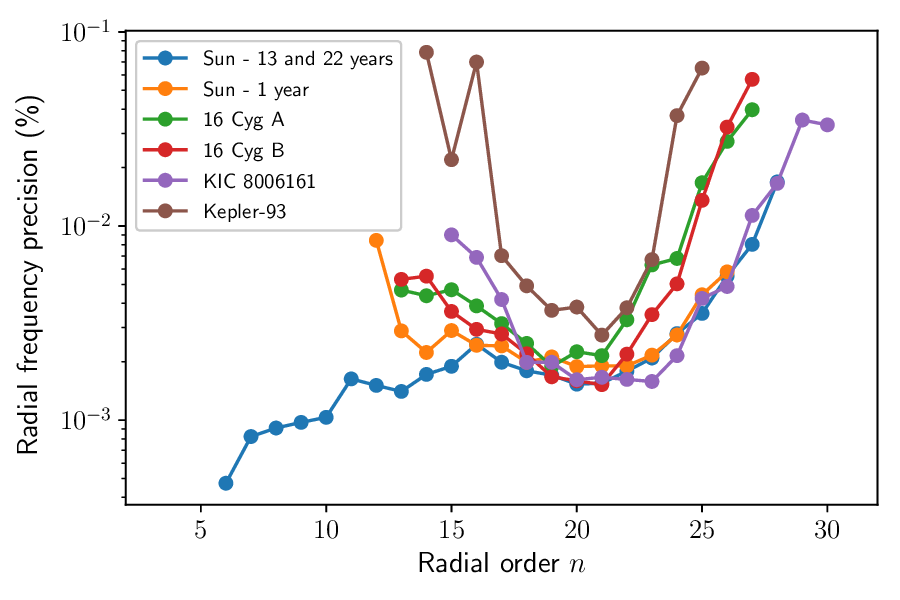} 
\caption{Comparison of observational precision of the frequencies of radial modes ($l=0$) between our mode sets, some of the best \textit{Kepler} observations (16~Cyg A and B, and KIC8006161), and a medium-quality \textit{Kepler} observation. The blue data correspond to set~0 and the orange data to set~2. Set~1 is identical to set~2 with the exception that the four lowest radial-order modes were removed.}
\label{fig:comp_freq_precision}
\end{figure}

For the consistency tests performed in Sect. \ref{sec:preliminary_tests}, we considered an additional very high-quality dataset, set~0. This set is based on the frequencies published by \citet{Basu2009} and supplemented by the frequencies of \citet{Davies2014} for the low-frequency modes $(n=6-13)$. The study by \citet{Basu2009} is based on approximately 13 years of continuous solar observations, while the study by \citet{Davies2014}, which focuses on low-frequency modes, is based on 22 years of data. In Table~\ref{tab:ranges_radial_orders}, we summarise the properties of the different mode sets considered in this study, and in Fig.~\ref{fig:comp_freq_precision}, we illustrate the observational precision of the frequencies for our mode sets, compared with some of the best \textit{Kepler} observations (16 Cyg A and B, and KIC8006161), and a medium-quality \textit{Kepler} observation (KIC3544595, and hereafter denoted as Kepler-93).

\begin{table}[t!]
\centering
\caption{Range of radial orders $n$ included in the different mode sets of this study.}
\begin{tabular}{lccc}
\hline \hline
Mode set & $l=0$ & $l=1$ & $l=2$ \\ 
\hline 
\textit{GOLF} &  &  &  \\ 
Set 0 & $6-28$ & $7-25$ & $8-26$ \\ 
Set 1 & $12-26$ & $12-26$ & $12-25$ \\ 
Set 2 & $16-26$ & $16-26$ & $16-25$ \\ 
\hline 
\textit{BiSON} &  &  &  \\ 
Set 1 & $12-25$ & $12-25$ & $12-25$ \\  
Set 2 & $16-25$ & $16-25$ & $16-25$ \\ 
\hline 
\end{tabular} 
\label{tab:ranges_radial_orders}
\end{table}

\subsection{Seismic inversions}
\label{sec:seismic_inversions}
In this section, we present a concise overview of the seismic inversions employed in our study. For a more comprehensive explanation of inversion techniques, we refer to \citet{Gough&Thompson1991}, \citet{Gough1993}, \citet{Pijpers2006}, \citet{Buldgen2022c}, and \citet{Betrisey2024_phd}. We adopt the terminology commonly used in the inversion community. The objective of an inversion is to determine the properties of an `observed' or `target' model, which can be either actual observational data -- in our case, from \citet{Betrisey2024_MA_Sun} -- or a synthetic target. To achieve this, the inversion uses a `reference' model as input, derived from a modelling procedure, typically the forward modelling, and linearly applies a small correction to a quantity of interest, known as a seismic indicator, calculated from the reference model. For clarity in this manuscript, we define an `inverted' quantity as the quantity of interest that includes the correction from the inversion. In our study, we use the forward models of \citet{Betrisey2024_MA_Sun} as reference models. We investigate the impact of magnetic activity on mean density and acoustic radius inversions. We focus on these two seismic indicators for two main reasons. Firstly, seismic inversions are integrated in the PLATO pipeline, therefore necessitating a thorough quantitative investigation of all systematic effects influencing them. Secondly, as demonstrated by \citet{Betrisey2024_AMS_quality}, mean density and acoustic radius inversions are suitable for large-scale application. More complex seismic indicators \citep[e.g.][]{Buldgen2015a,Buldgen2015b,Buldgen2018,Pijpers2021,Betrisey&Buldgen2022} are not, and their investigation would require a more advanced and computationally expensive modelling strategy. 

Seismic inversions rely on the structure inversion equation, which is grounded in the study of linear perturbations in stellar oscillations. \citet{Lynden-Bell&Ostriker1967}, along with earlier works by \citet{Chandrasekhar1964}, \citet{Chandrasekhar&Lebovitz1964}, and \citet{Clement1964}, established that the equation of motion adheres to a variational principle. Building on this finding, \citet{Dziembowski1990} demonstrated that, at first order, the perturbation in frequency can be directly linked to structural perturbations via the structure inversion equation:
\begin{equation}
\frac{\delta\nu^{n,l}}{\nu^{n,l}} = \int_{0}^{R} K_{\rho,\Gamma_1}^{n,l}\frac{\delta \rho}{\rho}dr + \int_{0}^{R} K_{\Gamma_1,\rho}^{n,l}\frac{\delta \Gamma_1}{\Gamma_1}dr + \mathcal{O}(\delta^2),
\label{eq:stucture_inversion_equation}
\end{equation}
where $\nu^{n,l}$ is the oscillation frequency of radial order $n$ and harmonic degree $l$, $K_{\rho,\Gamma_1}^{n,l}$ and $K_{\Gamma_1,\rho}^{n,l}$ are the structural kernels, $\rho$ is the density, and $\Gamma_1$ is the first adiabatic exponent. We note that $\Gamma_1=~\left(\frac{\partial\ln P}{\partial\ln \rho}\right)_{\mathrm{ad}}$, where $P$ is the pressure. Furthermore, we used the following definition:
\begin{equation}
\frac{\delta x}{x} = \frac{x_{\mathrm{obs}}-x_{\mathrm{ref}}}{x_{\mathrm{ref}}}.
\end{equation}
In this equation, `ref' and `obs' stand for reference and observed, respectively. Originally, \citet{Dziembowski1990} formulated this equation for the $(\rho, c^2)$ structural pair, where $c$ denotes the sound speed. However, the structure inversion equation is versatile and can be adapted for nearly any combination of physical variables present in the adiabatic oscillation equations \citep[see e.g.][]{Gough&Thompson1991,Gough1993,Elliott1996,Basu&JCD1997,Kosovichev1999,Lin&Dappen2005,Kosovichev2011,
Buldgen2015b,Buldgen2017a,Buldgen2018}. Based on the relative differences between observed and reference frequencies, a correction can then be computed on the mean density or the acoustic radius of the reference model by combining equations of the form given in Eq.~\eqref{eq:stucture_inversion_equation}. Given the limited number of modes in asteroseismology\footnote{up to 50 for the best \textit{Kepler} targets \citep[e.g.][]{Lund2017}. For comparison, higher-degree modes can be resolved in solar data and thousands of oscillation frequencies can therefore be extracted from the power spectrum \citep[e.g.][]{Larson&Schou2015,Reiter2020}, which allows us to localise the information and carry out profile inversions \citep[see review by][]{Buldgen2022c}.}, the objective is to define a global seismic indicator $t$, which encapsulates all the information from the frequency spectrum. In our context, this includes the mean density $\bar{\rho}$ and the acoustic radius $\tau$. We used the subtractive optimally localised averages \citep[SOLA;][]{Pijpers&Thompson1992,Pijpers&Thompson1994} method, where the following cost function is minimised:
\begin{align}
\mathcal{J}_{\bar{\rho}}(c_i) &= \int_0^1 \big(\mathcal{K}_{\mathrm{avg}} - \mathcal{T}_{t}\big)^2 dx 
                                        + \beta\int_0^1 \mathcal{K}_{\mathrm{cross}}^2dx + \lambda\left[k-\sum_i c_i\right] \nonumber\\
                                        &\quad+\tan\theta \frac{\sum_i (c_i\sigma_i)^2}{\langle\sigma^2\rangle} + \mathcal{F}_{\mathrm{Surf}}(\nu),
\label{eq:SOLA_cost_function}
\end{align}
The variable $x$ is defined as $x = r/R$, and $\mathcal{K}_{\text{avg}}$ and $\mathcal{K}_{\text{cross}}$ are the averaging and cross-term kernels, respectively. They are defined as follows:
\begin{align}
\mathcal{K}_{\mathrm{avg}} &= \sum_{i=1}^N c_i K_{\rho,\Gamma_1}^{i}\; , \\
\mathcal{K}_{\mathrm{cross}} &= \sum_{i=1}^N c_i K_{\Gamma_1,\rho}^{i}\; .
\end{align}
The aim of the SOLA approach is to achieve a good fit of the target function $\mathcal{T}_t$ while minimising the contributions from the cross-term and observational uncertainties. The variables $\theta$ and $\beta$ are trade-off parameters used to balance the different terms during minimisation, and $k$ is a normalisation constant depending on the properties of the indicator \citep[refer to][]{Buldgen2022c}. The inversion coefficients are denoted by $c_i$, with \( i \equiv (n, l) \) being the identification pair of an oscillation frequency, $\lambda$ is a Lagrange multiplier, $\langle\sigma^2\rangle = \sum_{i=1}^N\sigma_i^2$,  with $\sigma_i$ being the $1\sigma$ uncertainty of the relative frequency difference, and \( N \) is the total number of observed frequencies. The term $\mathcal{F}_{\text{Surf}}(\nu)$ provides an empirical description of the surface effects at the expense of introducing two additional free parameters into the minimisation (see Sect.~\ref{sec:treatment_SE_and_MA}).

For the mean density inversion, the target function is given by \citep{Reese2012}:
\begin{align}
\mathcal{T}_{\bar{\rho}}(x) =  4\pi x^2 \frac{\rho}{\rho_R},
\end{align}
where $\rho_R=M/R^3$, and $M$ and $R$ denotes the stellar mass and radius, respectively. For the acoustic radius inversion, the target function is defined as \citep{Buldgen2015a}:
\begin{align}
\mathcal{T}_{\tau}(x) = \frac{-1}{2\tau c}.
\end{align}
where $\tau = \int_0^1 \frac{dx}{c}$ is the acoustic radius. Following \citet{Reese2012} and \citet{Buldgen2015a}, we fixed the trade-off parameters to $\beta=10^{-6}$ and $\theta=10^{-2}$ for both inversions.

\subsection{Treatment of surface effects and magnetic activity}
\label{sec:treatment_SE_and_MA}
To evaluate the influence of the magnetic activity cycle on the asteroseismic characterisation, we use the same approach as in \citet{Betrisey2024_MA_Sun}. We assume a surface effect prescription and investigate whether the impact of magnetic activity can be masked within the free parameters of this parametrisation, as predicted by the literature \citep{Howe2017}. As pointed out by \citet{Salabert2018}, this indirect treatment of magnetic activity might not be effective if the frequency shifts due to magnetic activity do not increase monotonically with frequency as observed in the Sun. Furthermore, using a modelling strategy similar to that to be adopted in the PLATO pipeline, \citet{Betrisey2024_MA_Sun} demonstrated that this approach is insufficient to prevent an imprint of the activity cycle on the seismic age of the Sun. Among the three primary surface effect prescriptions in the literature \citep{Kjeldsen2008,Ball&Gizon2014,Sonoi2015}, we focus on the \citet{Ball&Gizon2014} prescription for the analysis in Sect.~\ref{sec:influence_magnetic_activity_cycle}. This prescription is widely adopted by the community and is considered the most robust \citep[see e.g. discussion in][]{Betrisey2023_AMS_surf}. In Sect.~\ref{sec:discussion}, we scrutinise this choice and provide a systematic uncertainty associated with the surface effect prescription.

In general, a seismic inversion is performed on an input model whose theoretical frequencies are not corrected for surface effects. In this study, we explored the two main methods to account for surface effects in the inversion procedure. The first method (labelled as `Inversion + coef. from AIMS' in the figures and tables) involves using the surface effect free parameters constrained by the forward modelling, therefore omitting the term $\mathcal{F}_{\text{Surf}}(\nu)$ in Eq.~\eqref{eq:SOLA_cost_function}. The second method (labelled as `Inversion + coef. from InversionKit' in the figures and tables) estimates these parameters within the inversion itself. Each approach has its merits and drawbacks. The latter is physically more consistent, as the forward modelling conducted by \citet{Betrisey2024_MA_Sun} with the Asteroseismic Inference on a Massive Scale (AIMS) software \citep{Rendle2019} interpolates within a grid of precomputed stellar models to find optimal stellar parameters but does not provide the corresponding stellar structure, which is a mandatory input for the inversion procedure. We note that a software\footnote{\url{https://lesia.obspm.fr/perso/daniel-reese/spaceinn/interpolatemodel/index.html}} was developed to interpolate the stellar structure based on the interpolation coefficients provided by AIMS. While this code produces a model in hydrostatic equilibrium by construction, there is no guarantee that the thermal equilibrium or the equation of state are strictly verified. We therefore decided to derive the stellar structure and associated adiabatic acoustic frequencies in a more robust way, respectively with the Liège evolution code \citep[CLES;][]{Scuflaire2008b} and the Liège oscillation code \citep[LOSC;][]{Scuflaire2008a}. For consistency, we employed the software versions used to generate the grid of stellar models. In practice, AIMS outputs two types of variables: the optimised free variables (in our case, mass, age, hydrogen and helium initial mass fractions, and surface effect coefficients) and the associated stellar parameters (mean density, effective temperature, absolute luminosity, etc.). The optimal stellar structure is recomputed using the free parameters from the forward modelling. This procedure yields a stellar structure very close to the one corresponding to the optimal parameters from the forward modelling but not exactly matching the associated stellar parameters. For instance, the mean densities of the recomputed structure and those from AIMS are slightly different (see Sect.~\ref{sec:capabilities_seismic_inversions}). This discrepancy is not problematic for the inversion, which will provide a slightly larger correction if necessary. However, the surface effect parameters are not fully consistent with the recomputed structure. The differences between the associated stellar parameters of AIMS and the recomputed structure are much smaller than other sources of uncertainties, making it reasonable to neglect them. In the second approach, the free parameters of the surface effect prescription are consistently estimated within the inversion, introducing two additional free parameters into the minimisation. Consequently, the inversion becomes numerically less stable, as illustrated for example in \citet{Betrisey2024_AMS_quality}. The optimal approach depends on the conditions under which the inversion is performed. As discussed in more detail in Sect.~\ref{sec:discussion}, based on the results of \citet{Betrisey2023_AMS_surf}, \citet{Betrisey2024_AMS_quality}, and this study, the differences in inversion outcomes for mean density and acoustic radius between the two methods are negligible compared to other sources of uncertainty. For a typical asteroseismic target, the first approach is preferred, while for exceptional data quality, such as in this study with solar data, the second approach could also be utilised.

\clearpage

\section{Influence of the magnetic activity cycle}
\label{sec:influence_magnetic_activity_cycle}
\begin{figure*}[p!]
\centering
\resizebox{0.91\linewidth}{!}{
\begin{subfigure}[b]{.49\textwidth}
  \includegraphics[width=.99\linewidth]{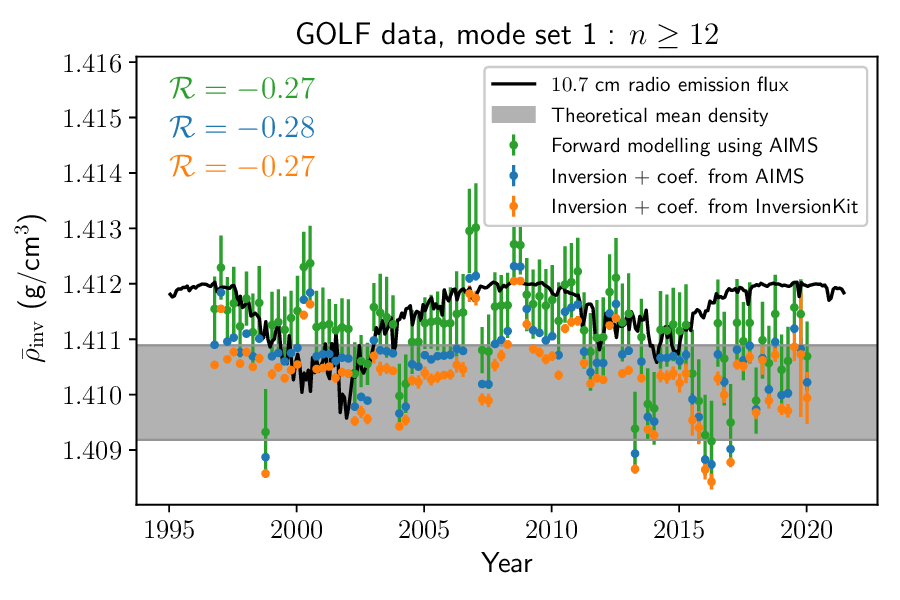}  
\end{subfigure}
\begin{subfigure}[b]{.49\textwidth}
  \includegraphics[width=.99\linewidth]{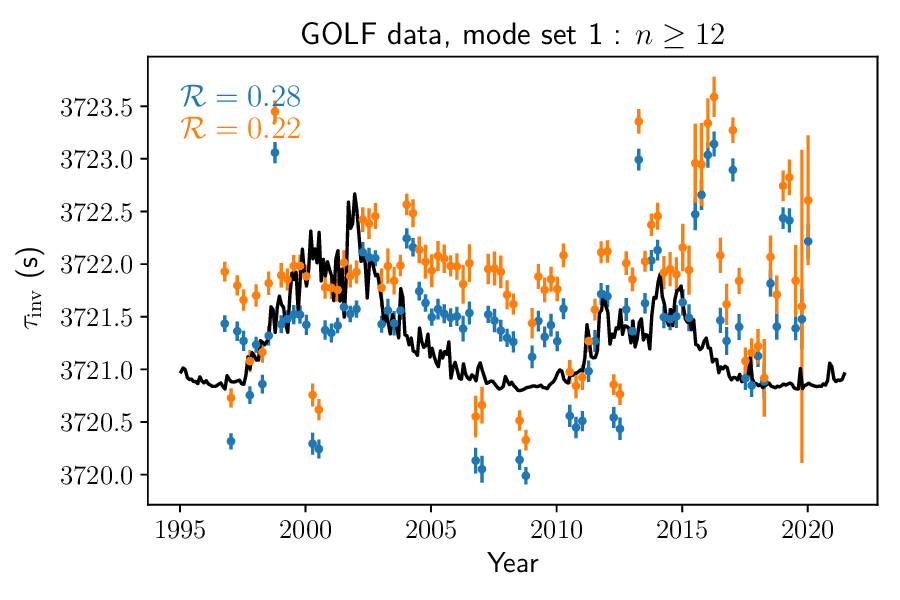} 
\end{subfigure}}
\resizebox{0.91\linewidth}{!}{
\begin{subfigure}[b]{.49\textwidth}
  \includegraphics[width=.99\linewidth]{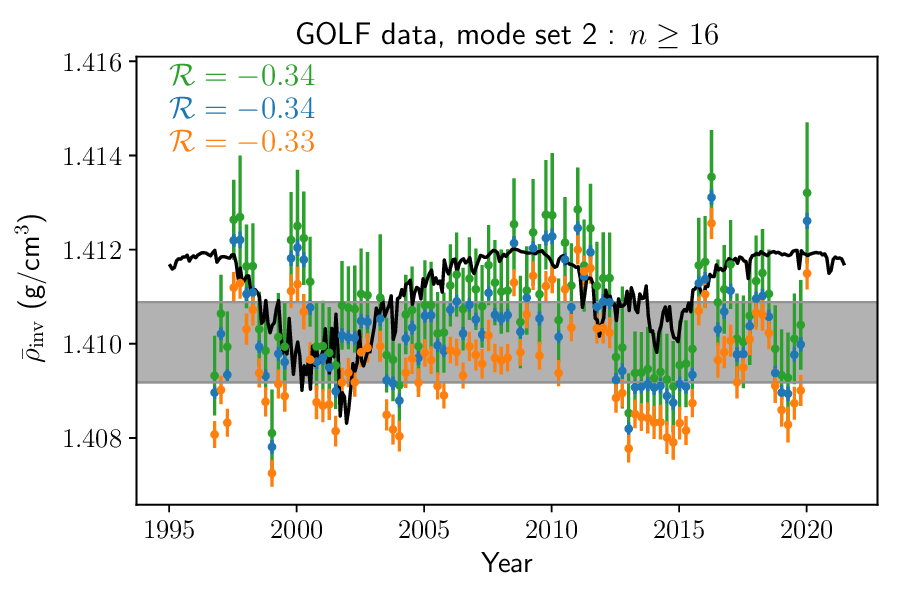}  
\end{subfigure}
\begin{subfigure}[b]{.49\textwidth}
  \includegraphics[width=.99\linewidth]{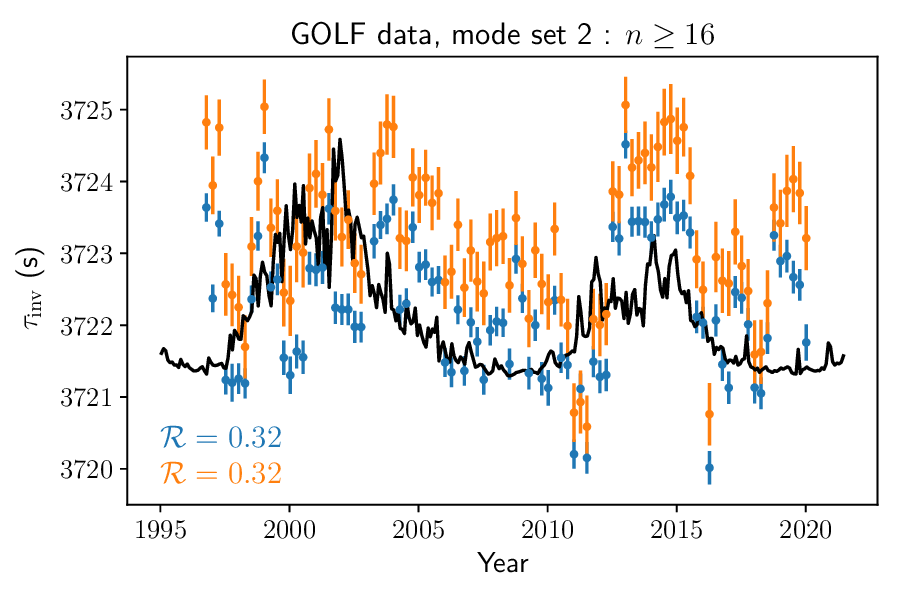} 
\end{subfigure}}
\resizebox{0.91\linewidth}{!}{
\begin{subfigure}[b]{.49\textwidth}
  \includegraphics[width=.99\linewidth]{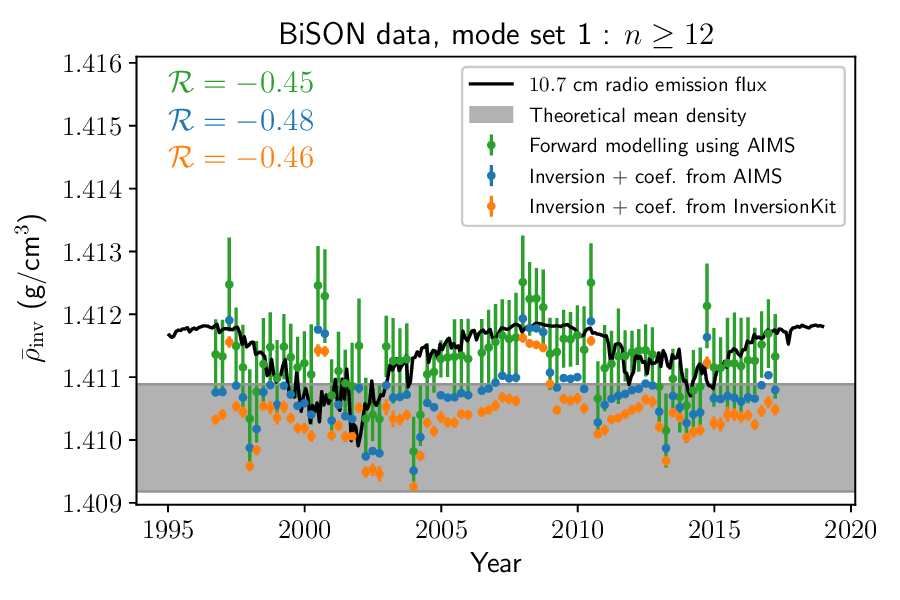}  
\end{subfigure}
\begin{subfigure}[b]{.49\textwidth}
  \includegraphics[width=.99\linewidth]{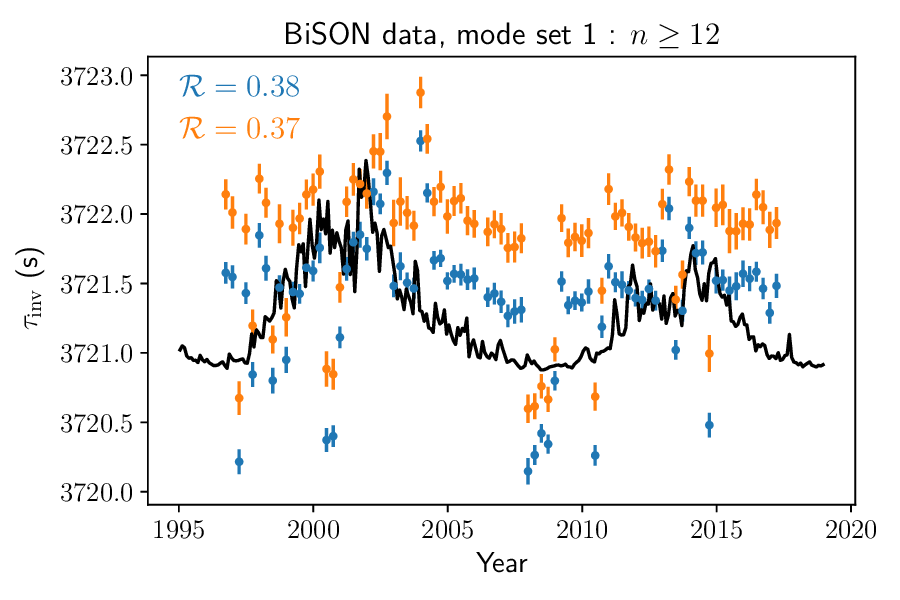} 
\end{subfigure}}
\resizebox{0.91\linewidth}{!}{
\begin{subfigure}[b]{.49\textwidth}
  \includegraphics[width=.99\linewidth]{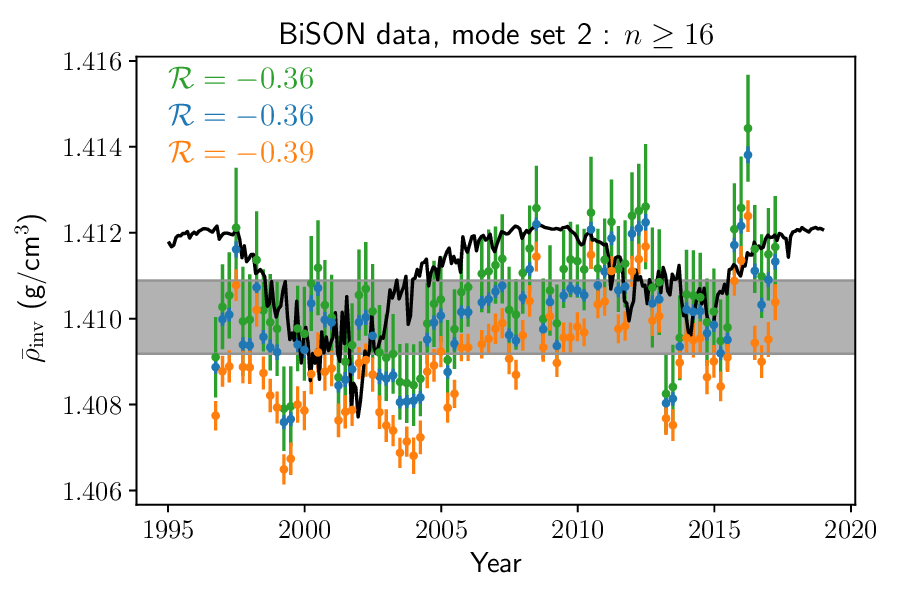}  
\end{subfigure}
\begin{subfigure}[b]{.49\textwidth}
  \includegraphics[width=.99\linewidth]{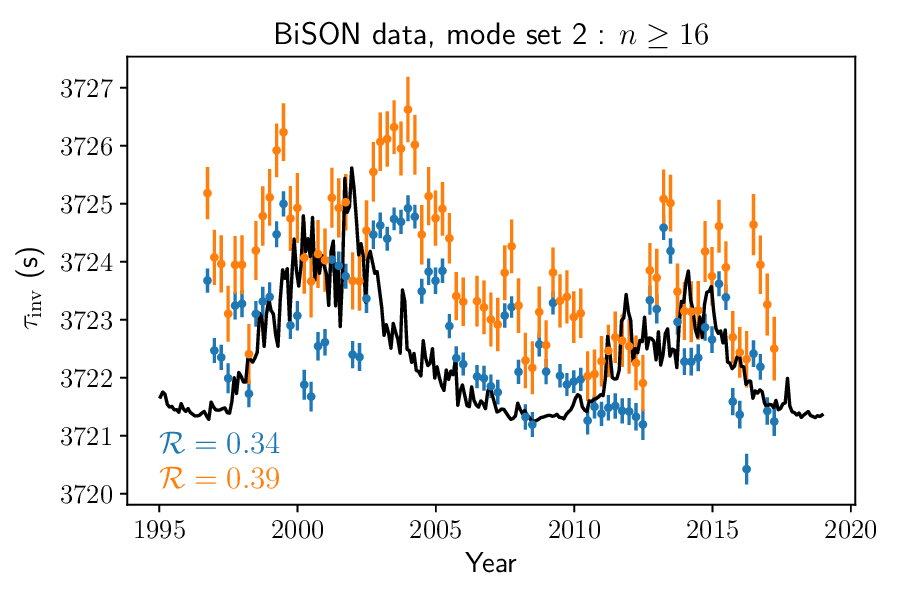} 
\end{subfigure}}
\caption{Impact of the magnetic activity cycle on mean density (left column) and acoustic radius (right column) inversions based on GOLF (upper half) and BiSON (lower half) data. The results from the forward modelled are displayed in green. The forward modelling results are shown in green, while the inversion results are illustrated in blue and orange, depending on whether the surface effect coefficients are derived from AIMS or InversionKit. The activity proxy, the 10.7 cm radio emission flux, is represented by the solid black line. For illustration purposes, this proxy was rescaled and reversed in cases of negative correlation (left column). The gray band is the $1\sigma$ interval for the theoretical mean density computed from the solar total mass $M_\odot$ and radius $R_\odot$. The number $\mathcal{R}$ denotes the Spearman rank correlation coefficient between the activity proxy and the corresponding dataset.}
\label{fig:results_golf}
\end{figure*}

The inversions in this study were performed using the InversionKit software \citep{Reese&Zharkov2016}, which was initially developed for individual modelling. Building on the works of \citet{Betrisey2022,Betrisey2023_AMS_surf,Betrisey2024_AMS_quality,
Betrisey2024_MA_Sun} to automate the Forward and Inversion COmbination (FICO) procedure -- a sophisticated modelling strategy that integrates forward and inverse methods to mitigate surface effects and yield precise and accurate stellar parameters \citep[refer to Sect.~3.3 of][and references therein for a detailed explanation of the modelling strategy]{Betrisey2024_phd} -- we have modified the InversionKit software to support large-scale automated applications. To verify the effectiveness of this automation, we performed consistency tests detailed in Appendix~\ref{sec:preliminary_tests}. In this section, we present the results for the GOLF and BiSON datasets.

The correlation between the activity proxy and the different stellar parameters was estimated using a similar approach to that of \citet{Betrisey2024_MA_Sun}, with one key modification. Instead of calculating the Pearson correlation coefficient \citep{Pearson1895}, we employed the Spearman correlation coefficient \citep{Spearman1904}. In their study, \citet{Betrisey2024_MA_Sun} assumed an approximately linear relationship between the activity proxy and the different stellar parameters they examined. We did not make this assumption and chose to use the Spearman correlation coefficient instead. By comparing our Spearman coefficients with the results of \citet{Betrisey2024_MA_Sun}, we found that our correlation coefficients are systematically higher, suggesting that the relationship is likely not entirely linear. However, the differences are minor, and using the Spearman coefficients in \citet{Betrisey2024_MA_Sun} would not alter their conclusions. On the contrary, it would reinforce them, as they observed strong correlations for some stellar parameters, notably the solar age, using a suboptimal measure.

Additionally, to evaluate the accuracy of the mean density inversion results, we derived a theoretical estimate based on fundamental observables measured independently from helioseismology. While the robustness of mean density inversion is well established in the literature \citep[e.g.][]{Reese2012,Buldgen2015a}, it remains valuable to reaffirm this with our study. The mean density can be determined using the solar radius and total mass, which is itself derived from the solar mass parameter and the Newtonian gravitational constant:
\begin{align}
\bar{\rho}_{\rm theo} = \frac{(GM)_\odot}{\frac{4\pi G}{3}R_\odot^3},
\label{eq:rho_theo}
\end{align}
where $(GM)_\odot = (1.3271244\pm 0.0000001)\cdot 10^{26}$ cm$^3$/s$^2$ is the solar mass parameter \citep[IAU 2015, Resolution B3;][]{IAU2015-ResolutionB3}, $G = (6.6743015\pm 0.0001468)\cdot 10^{-8}$ cm$^3$/g/s$^2$ is the Newtonian gravitational constant\footnote{\citet{CODATA2018} provide a relative uncertainty on $G$, namely $\sigma_G / G = 2.2\cdot 10^{-5}$.} \citep[CODATA 2018;][]{CODATA2018}, and $R_\odot = (6.95658\pm 0.00140)\cdot 10^{10}$ cm is the solar radius\footnote{The recommended nominal value of $R_\odot^N = 6.957\cdot 10^8$ m by the IAU 2015 Resolution B3 is the rounded value of \citet{Haberreiter2008} by keeping only one significant digit for the uncertainty \citep[see][]{IAU2015-ResolutionB3}.} \citep{Haberreiter2008}. As discussed for example in \citet{Haberreiter2008} and \citet{Takata&Gough2024}, the solar radius in not unequivocally defined and we therefore decided to base our theoretical mean density estimate on widely-used values. Similarly, the solar mass parameter as defined by \citet{IAU2015-ResolutionB3} is a nominal value utilised for consistent unit conversions in the literature. Consequently, the authors do not specify an uncertainty. According to the explanations provided in Resolution B3 and by \citet{Luzum2011}, the number of significant digits was selected to ensure that the barycentric coordinate and dynamical times align within this precision. Thus, we adopted an uncertainty that aligns with this rationale. It is important to note that this method yields an extremely conservative uncertainty, as the precision achieved for both barycentric times is three orders of magnitude smaller than our assumption \citep{Folkner2009}. As an aside, we note that the precision of the solar mass parameter is much higher than the other parameters in Eq.~\eqref{eq:rho_theo}, even with our conservative estimate, and we could have assumed it to be exact for simplicity. As a result, we adopt 
\begin{align}
\bar{\rho}_{\rm theo} = (1.4100\pm 0.0009)\ \mathrm{g/cm}^3
\end{align}
in our study. We note that the quoted uncertainty is fully dominated by the uncertainty on the solar radius. As the acoustic radius is computed by integrating the inverse of the sound speed profile, it is not possible to constrain this quantity only based on fundamental stellar observables as we did for the mean density.

\citet{Betrisey2024_MA_Sun} observed a weak imprint on the solar mean density as estimated by forward modelling, using a strategy akin to that planned for the PLATO pipeline. As discussed by the authors, this imprint is an indirect consequence of the impact of magnetic activity on other stellar parameters, particularly stellar age, since the mean density was not included as a free variable in the minimisation process. In the left column of Fig.~\ref{fig:results_golf}, we present the temporal evolution of the mean density over two solar cycles, as estimated by seismic inversions. As outlined in Sect.~\ref{sec:treatment_SE_and_MA}, surface effects were modelled using two methods, shown in blue and orange. For comparison, the forward modelling results are depicted in green. Both methods reveal a weak (and direct) imprint of the magnetic activity cycle on the mean density, with Spearman coefficients around 0.3 with GOLF data and 0.4 with BiSON data. This observation was confirmed by smoothing the data with a Savitzky-Golay filter \citep{Savitzky&Golay1964}, which allowed us to visually identify two distinct peaks corresponding to the cycle maxima. Although the smoothing is not as clean as in observations by \citet{Betrisey2024_MA_Sun} of solar age due to a weaker correlation, the signal remains significant. The imprint is comparable to that found by forward modelling, as indicated by the similar Spearman correlation coefficients.

Consistent with the findings of \citet{Betrisey2024_MA_Sun}, we observe that including low radial-order modes in the datasets mitigates the effects of magnetic activity. This finding is expected as low radial-order oscillations in the Sun are less sensitive to the effects of magnetic activity. This mitigation is however more significant that on the imprint on stellar age obtained by \citet{Betrisey2024_MA_Sun} through forward modelling. On average, the imprint is reduced from 0.15\%\footnote{This number corresponds the systematic uncertainty due to magnetic activity, expressed as a percentage of the average value of the underlying dataset.} to 0.10\% and from 0.19\% to 0.08\% for GOLF and BiSON data, respectively (see Sect.~\ref{sec:discussion} for the derivation of these numbers).

Regarding the acoustic radius inversions, our findings are similar to those for mean density, as illustrated in the right column of Fig.~\ref{fig:results_golf}. We detect a weak imprint of the magnetic activity cycle, with Spearman correlation coefficients around 0.3 with GOLF data and 0.4 with BiSON data. For datasets that include low radial-order modes, the effectiveness of the inversion diminishes starting around 2015. This trend is also observed in the corresponding mean density inversions, although the mean density itself is less affected, with differences primarily reflected in the error bars. This behaviour is attributed to the ageing of the GOLF instrument \citep{Garcia2005}, which results in less precise frequency measurements, especially at low radial orders (see Appendix~\ref{app:supplementary_data}). This assumption is supported by the absence of this trend in BiSON data, whose ground-based observing facilities allow for better maintenance of the instruments over the years. Introducing low-order radial modes also significantly mitigates the effects of magnetic activity, from 0.026\%\footnotemark[4] to 0.017\% and from 0.033\% to 0.016\% for GOLF and BiSON data, respectively. 

We observe minor but systematic discrepancies between the green, blue, and orange data in Fig.~\ref{fig:results_golf}. The green data, sourced from \citet{Betrisey2024_MA_Sun}, employed a forward modelling algorithm, which differs from the seismic inversion techniques utilised in this study, leading to the observed offset. The discrepancy between the blue and orange data is attributed to the surface effect prescription. As detailed in Sect.~\ref{sec:treatment_SE_and_MA}, we applied two distinct methods to derive the surface effect coefficients, resulting in a slight offset. Additionally, since the blue data is based on surface effect coefficients from the forward modelling, it logically falls between the green and orange data. As discussed in Sect.~\ref{sec:discussion}, these offsets are negligible compared to other sources of uncertainty. Similarly, although the data at solar minima appears to align more closely with the theoretical mean density estimate, coherently from a physical standpoint, as the activity effects shifting the oscillation frequencies are minimal during these periods, it is challenging to consider this finding significant given the other sources of systematic uncertainty.

\section{Discussion}
\label{sec:discussion}
In this section, we first quantify the various sources of uncertainties affecting the inversion results, including statistical uncertainty, the selection of physical ingredients in the stellar models, the choice of surface effect prescription, and the influence of magnetic activity. Following this, we evaluate the performance of the inversions conducted in this study in Sect.~\ref{sec:capabilities_seismic_inversions}, and outline best practices and recommended values in Sect.~\ref{sec:recommended_values_best_practices}.

\subsection{Systematic uncertainties}
\label{sec:systematic_uncertainties}
\begin{table}[t!]
\centering
\caption{Relative uncertainty budget of the inverted mean density and acoustic radius.}
\begin{tabular}{lcc}
\hline \hline
 & $\bar{\rho}_{\mathrm{inv}}$ & $\tau_{\mathrm{inv}}$ \\ 
\hline 
\textit{Statistical uncertainty} & & \\
Inversion + coef. from AIMS & $\lesssim 0.01\%$ & $\lesssim 0.006\%$ \\ 
Inversion + coef. from InversionKit & $\lesssim 0.02\%$ & $\lesssim 0.013\%$ \\
\hline
 \textit{Systematic uncertainties} & & \\
Stellar activity & 0.19\% & 0.067\% \\ 
Surface effect prescription & 0.03\% & 0.012\% \\ 
Physical ingredients & 0.13\% & 0.026\% \\ 
\hline
\textit{Total uncertainty} & & \\
 & 0.36\% & 0.11\% \\ 
\hline 
\end{tabular} 
\label{tab:all_uncertainties}
\end{table}

\begin{table*}[t]
\centering
\caption{Quantitative assessment of the impact of magnetic activity on the different configurations investigated in this study.}
\resizebox{\linewidth}{!}{
\begin{tabular}{lcccccccc}
\hline \hline
\multirow{3}{*}{Specificities of the modelling procedure} & \multicolumn{4}{c}{$\bar{\rho}_{\mathrm{inv}}$} & \multicolumn{4}{c}{$\tau_{\mathrm{inv}}$} \\ \cmidrule[0.4pt](lr){2-5} \cmidrule[0.4pt](l){6-9}
 & \multicolumn{2}{c}{GOLF} & \multicolumn{2}{c}{BiSON} & \multicolumn{2}{c}{GOLF} & \multicolumn{2}{c}{BiSON} \\ \cmidrule[0.4pt](lr){2-3} \cmidrule[0.4pt](lr){4-5} \cmidrule[0.4pt](lr){6-7} \cmidrule[0.4pt](l){8-9} 
 & std & min/max & std & min/max & std & min/max & std & min/max \\
\hline
$n \geq 12$ \\ 
Forward modelling using AIMS & 0.05\% & 0.27\% & 0.05\% & 0.29\% & - & - & - & - \\  
Inversion + coef. from AIMS & 0.05\% & 0.25\% & 0.04\% & 0.28\% & 0.02\% & 0.08\% & 0.02\% & 0.13\% \\ 
Inversion + coef. from InversionKit & 0.05\% & 0.26\% & 0.04\% & 0.28\% & 0.02\% & 0.09\% & 0.02\% & 0.14\% \\ 
\hline 
$n \geq 16$ \\
Forward modelling using AIMS & 0.08\% & 0.39\% & 0.10\% & 0.76\% & - & - & - & - \\ 
Inversion + coef. from AIMS & 0.07\% & 0.38\% & 0.09\% & 0.72\% & 0.03\% & 0.12\% & 0.03\% & 0.23\% \\ 
Inversion + coef. from InversionKit & 0.08\% & 0.38\% & 0.10\% & 0.70\% & 0.03\% & 0.12\% & 0.03\% & 0.22\% \\ 
\hline 
\end{tabular}}
{\par\small\justify\textbf{Notes.} The `std' column represents the standard deviation of the data, expressed as a percentage of the mean value. The `min/max' column shows the difference between the maximum and minimum values, also expressed as a percentage of the mean value of the dataset. \par}
\label{tab:results_percentages_main}
\end{table*}

Following standard OLA practices, the statistical uncertainty is derived by propagating the observational uncertainty of the oscillation frequencies into the inversion result \citep[see e.g.][]{Backus&Gilbert1970,Pijpers&Thompson1994,Reese2012,Buldgen2022c}. For instance, in the mean density inversion, the observational uncertainties are propagated as described by Reese et al. (2012):
\begin{align}
\sigma_{\bar{\rho}_{\mathrm{inv}}} = s\cdot\bar{\rho}_{\mathrm{ref}}\left(\sum_{i=1}^N c_i^2 \sigma_i^2\right)^{1/2}.
\label{eq:statistical_uncertainty}
\end{align}
Table~\ref{tab:all_uncertainties} presents the upper bounds of the statistical uncertainties for the inverted mean density and acoustic radius. These bounds represent the average statistical uncertainties of each subseries of GOLF and BiSON data for set~2. If additional lower radial-order modes can be detected, the statistical uncertainty will decrease, which is why these values are considered upper bounds. For example, set~1, which includes four additional modes, shows an improvement in uncertainty by a factor of two to three. Additionally, we observe significantly larger statistical uncertainties when the surface effect coefficients are estimated within InversionKit. This result aligns with our expectations, as estimating these coefficients in InversionKit introduces additional free parameters in the minimisation, naturally increasing the statistical uncertainty. As noted by \citet{Reese2012}, there are other sources of uncertainty beyond those given in Eq. \eqref{eq:statistical_uncertainty}. Key sources of systematic uncertainties include the physical ingredients used in stellar models \citep[e.g.][]{Buldgen2019f,Farnir2020,Betrisey2022}, surface effects \citep[e.g.][]{Ball&Gizon2017,Nsamba2018,Jorgensen2020,Jorgensen2021,Cunha2021,Betrisey2023_AMS_surf}, and stellar magnetic activity \citep[e.g.][]{Broomhall2011,Santos2018,PerezHernandez2019,Santos2019_sig,Santos2019_rot,Howe2020,Thomas2021,Santos2021,Santos2023,
Betrisey2024_MA_Sun}.

Based on the findings in Sect.~\ref{sec:influence_magnetic_activity_cycle}, we can estimate the systematic uncertainty attributed to magnetic activity. For each dataset, we consider two error measures: the weighted standard deviation and half the difference between the maximum and minimum values. These values are summarised in Table~\ref{tab:results_percentages_main} for each dataset. These two error measures provide a lower and upper limit of the systematic uncertainty, respectively. The latter measure is particularly sensitive to data, especially if there are outlying datapoints, and tends to overestimate the uncertainty. Since the impact of magnetic activity on the inverted mean density and acoustic radius is correlated with the solar cycle, it is not a stochastic phenomenon. Therefore, using the standard deviation of the data will not meaningfully capture the amplitude of the variation during a cycle and will underestimate the uncertainty. However, these two measures provide a range within which the systematic uncertainty should fall. As a compromise, we estimate the systematic uncertainty due to magnetic activity as twice the largest standard deviation of the different datasets considered in this study. This conservative approach has the advantage of encompassing most of the variation during a cycle and not being sensitive to outliers. As shown in Table~\ref{tab:all_uncertainties}, the relative systematic uncertainty is 0.19\% for the inverted mean density and 0.067\% for the acoustic radius. Since both values are significantly larger than the statistical uncertainty, we conclude that the effects of magnetic activity cannot be ignored for these stellar parameters. Similarly to the results of Sect.~\ref{sec:influence_magnetic_activity_cycle}, we also find that the introduction of low radial-order modes can significantly mitigate the influence of magnetic activity. This mitigation effect is physically expected as low-frequency modes do not experience much of a solar cycle variation \citep[see e.g.][for a review]{Broomhall&Nakariakov2015}. Looking beyond solar characterisation, it should be noted that including low radial-order modes might lessen the impact of magnetic activity in some cases, but it does not constitute a universal solution as other stars may not see the same frequency dependence on the frequency shifts as the Sun \citep{Salabert2018}.

\begin{figure*}[t!]
\resizebox{0.99\linewidth}{!}{
\begin{subfigure}[b]{.49\textwidth}
  \includegraphics[width=.99\linewidth]{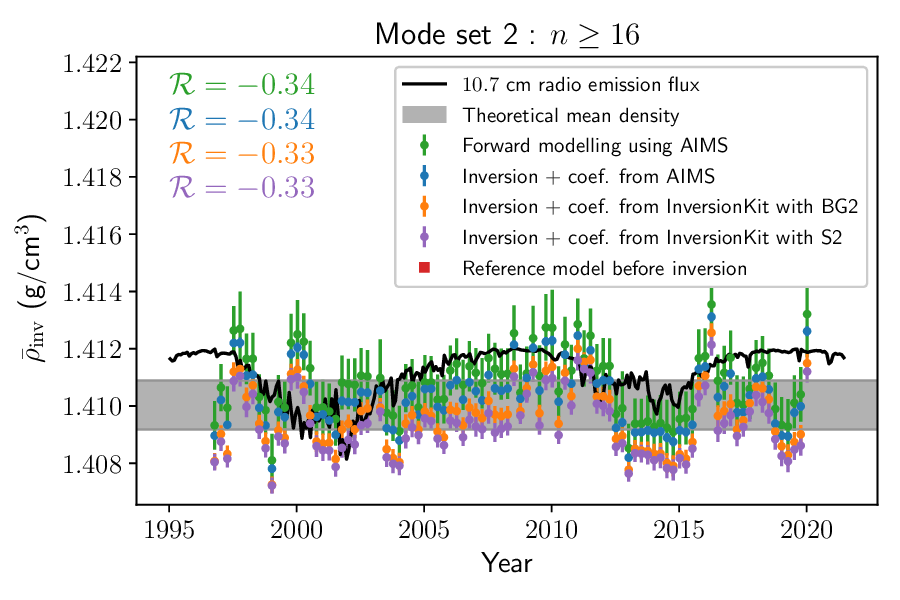}  
\end{subfigure}
\begin{subfigure}[b]{.49\textwidth}
  \includegraphics[width=.99\linewidth]{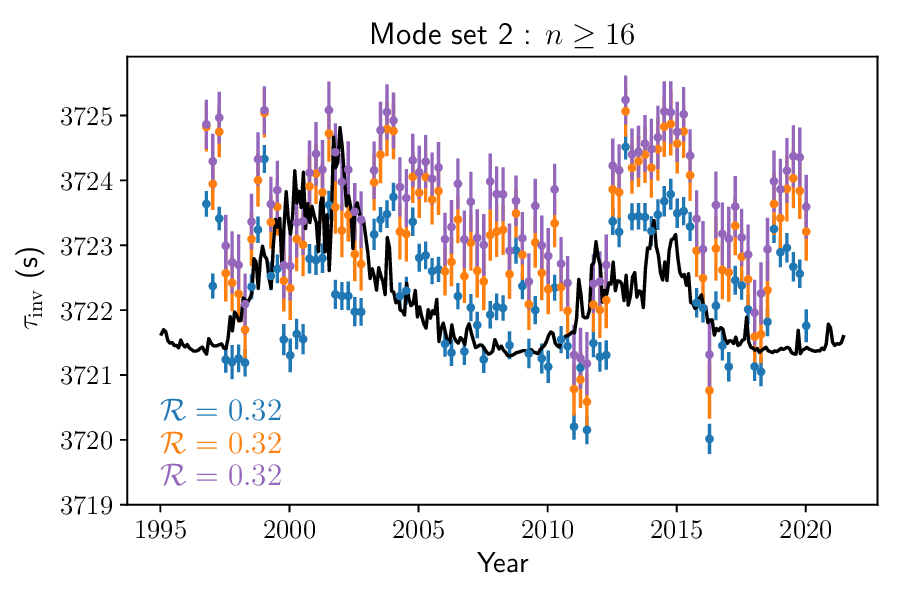} 
\end{subfigure}}
\resizebox{0.99\linewidth}{!}{
\begin{subfigure}[b]{.49\textwidth}
  \includegraphics[width=.99\linewidth]{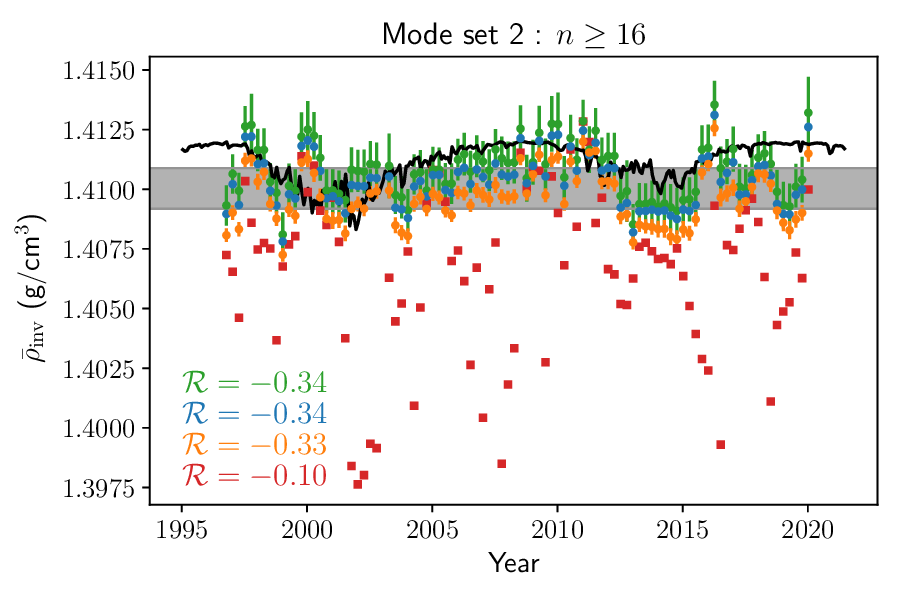}  
\end{subfigure}
\begin{subfigure}[b]{.49\textwidth}
  \includegraphics[width=.99\linewidth]{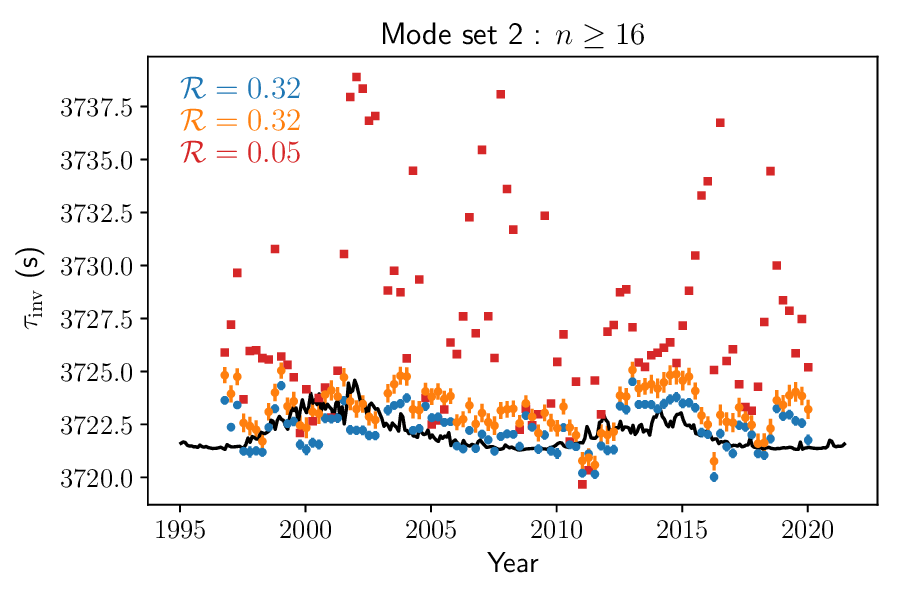} 
\end{subfigure}}
\caption{\textit{Upper panels:} Impact of surface effect prescription on mean density (left column) and acoustic radius (right column) inversions.  \textit{Lower panels:} Mean density (left column) and acoustic radius (right column) inversion performances. These tests are based on mode set 2 with GOLF data. The activity proxy, the 10.7 cm radio emission flux, is represented by the solid black line. It was rescaled for illustration purposes and reversed in the event of a negative correlation (left column). The gray band is the $1\sigma$ interval for the theoretical mean density computed from the solar total mass $M_\odot$ and radius $R_\odot$. The number $\mathcal{R}$ denotes the Spearman correlation coefficient between the activity proxy and the corresponding dataset.}
\label{fig:discussion}
\end{figure*}

To assess the systematic uncertainty linked to the choice of the surface effect prescription, we re-performed the seismic inversions on set~2 using GOLF data, this time changing the prescription to the \citet{Sonoi2015} correction. The upper panels of Fig.~\ref{fig:discussion} illustrate a comparison between the inversion results using the \citet{Ball&Gizon2014} and \citet{Sonoi2015} prescriptions. We concentrated on this set due to its lowest data quality, which results in the largest variations. The relative uncertainty was calculated by determining the relative difference between the weighted means of the data based on the \citet{Sonoi2015} and \citet{Ball&Gizon2014} corrections that we provided in Table~\ref{tab:recommended_values} for completeness. Given the unreliability of the \citet{Kjeldsen2008} prescription \citep[e.g.][]{Ball&Gizon2017,Nsamba2018,Betrisey2023_AMS_surf}, it was not included in our analysis. We identified relative systematic uncertainties of 0.026\% for the inverted mean density and 0.012\% for the acoustic radius. These uncertainties should be interpreted with caution, as our analysis utilised solar data, and the surface effect prescriptions are optimised for this context. When applied to non-solar data, particularly for stars with convective cores, the discrepancies can become significantly larger \citep[see e.g.][]{Betrisey2023_AMS_surf}. The systematic uncertainty due to the choice of the physical ingredients was estimated in \citet{Betrisey2024_phd} by considering a comprehensive set of physical ingredients and conducting local minimisations to deduct the contribution of the different physics.

\begin{table*}[t]
\centering
\caption{Theoretical expectations, seismic values of in-depth investigations of mode set 2 based on GOLF data, and recommended values of solar mean density and acoustic radius as determined through seismic inversions.}
\begin{tabular}{lcc}
\hline \hline
 & $\bar{\rho}$ (g/cm$^3$) & $\tau$ (s) \\ 
\hline
\textit{Theoretical expectations} & & \\
 & $1.4100\pm 0.0009$ & - \\ 
\hline
 \textit{Seismic values (set 2 of GOLF)} & & \\
Forward modelling  & $1.4107\pm 0.0011$ & - \\ 
Reference model before inversion & $1.4066\pm 0.0035$ & $3727.3\pm 4.4$ \\ 
Inversion + coef. from AIMS with BG2 & $1.4103\pm 0.0011$ & $3722.3\pm 0.9$ \\ 
Inversion + coef. from InversionKit with BG2 & $1.4096\pm 0.0011$ & $3723.2\pm 1.0$ \\  
Inversion + coef. from InversionKit with S2 & $1.4093\pm 0.0010$ & $3723.7\pm 0.9$ \\ 
\hline
\textit{Recommended values} & & \\
  & $1.4104\pm 0.0051$ & $3722.0\pm 4.1$ \\ 
\hline 
\end{tabular}
{\par\small\justify\textbf{Notes.} The recommended values correspond to the weighted average of the inversion results using the surface effect coefficients estimated by the forward modelling with AIMS. The average is based on results from mode sets 1 and 2 of GOLF and BiSON datasets. The uncertainty of the seismic values was estimated using the standard deviation of the corresponding data and therefore only accounts for half of the systematic uncertainty due to stellar activity. The uncertainty of the recommended values accounts for all the uncertainties listed in Table \ref{tab:all_uncertainties} (i.e. statistical, stellar activity, surface effect prescription, and physical ingredients). \par}
\label{tab:recommended_values}
\end{table*}

\subsection{Seismic inversion performances}
\label{sec:capabilities_seismic_inversions}
In the lower panels of Fig.~\ref{fig:discussion}, we thoroughly examined the performance of the seismic inversions that were employed in this study, focusing on mode set 2 of GOLF data. The numerical values are detailed in Table~\ref{tab:recommended_values}. The figure shows that the differences between inversions based on surface effect coefficients estimated by AIMS (blue data) or directly within the inversion (orange data) are negligible in the solar case. Specifically, an absolute difference of 0.0007 g/cm$^3$ was observed between the two methods. This difference is smaller than the statistical uncertainty, which is already negligible compared to the other sources of uncertainties discussed in the previous section.

Although the robustness of mean density and acoustic radius inversions has been extensively demonstrated in the literature \citep[e.g.][]{Reese2012,Buldgen2015a}, the lower panels of Fig.~\ref{fig:discussion} further illustrate this robustness. As detailed in Sect.~\ref{sec:treatment_SE_and_MA}, AIMS does not compute the interpolated stellar structure, which is a necessary input for seismic inversions. While it is straightforward to recompute this structure a posteriori based on the optimal stellar parameters estimated by AIMS, the mean density and acoustic radius derived in this manner (red data) are suboptimal compared to the AIMS results (green data). However, seismic inversions can fully compensate for this and yield inverted quantities (blue or orange data) that are more precise and accurate than the AIMS forward modelling. Additionally, we note that some reference models used for the inversion are above the inversion convergence region, and we verified that the correction is indeed in the opposite direction compared to the models below the convergence region. In the solar case, forward models are already of high quality, and the improvement with seismic inversions confirms the quality of the forward modelling rather than providing a significant quantitative gain, unlike in the case of solar-like stars, which are more challenging to accurately reproduce with forward modelling alone. As an aside, it should be noted that although the gain in accuracy is negligible compared to other sources of uncertainty in the solar case, a small improvement was still consistently observed in most of the configurations that were tested.

\subsection{Recommended values and best practices}
\label{sec:recommended_values_best_practices}
In Table~\ref{tab:recommended_values}, we present the recommended values for the solar mean density and acoustic radius, determined through seismic inversions and based on observations spanning two full solar cycles by GOLF and BiSON. These recommended values are the averages of results from mode sets 1 and 2 of both datasets. The suggested uncertainties encompass all the factors listed in Table~\ref{tab:all_uncertainties}, including statistical uncertainty, and stellar activity, surface effect prescription, and the systematics related to physical ingredients.

Additionally, these values are derived from the inverted results using the surface effect coefficients estimated by AIMS forward modelling. It is worth noting that, in the specific case of the Sun, we could have employed the alternative approach of estimating the surface effect coefficients directly within the inversion. However, this method is numerically less stable \citep[see e.g.][]{Betrisey2024_AMS_quality}, particularly for typical asteroseismic targets. Consequently, we obtained final recommended values of $\bar{\rho}_{\mathrm{inv}} = 1.4104\pm 0.0051$ g/cm$^3$ and $\tau_{\mathrm{inv}} = 3722.0\pm 4.1$ s. The mean density aligns well with theoretical predictions based on the recommended values by the IAU 2015 Resolution B3 for solar mass and radius. Furthermore, this value is consistent with the inverted mean density derived from the data presented by \citet[][hereafter B09]{Broomhall2009_def}, taking into account the raw frequency set ($\bar{\rho}_{\mathrm{inv,raw}}^{\mathrm{B09}}= 1.4089\pm 0.0051$) or the set adjusted for solar cycle effects ($\bar{\rho}_{\mathrm{inv,adjusted}}^{\mathrm{B09}}= 1.4092\pm 0.0051$). Generally, caution is advised when comparing values from different datasets due to potential offsets arising from data selection and processing methods, such as frequency readjustments for solar cycle effects. However, in our study, we compare values derived from decades of observations and quasi-complete solar cycles. Thus, the offsets are negligible relative to other sources of uncertainty, leading to the high degree of consistency that we observed.

\section{Conclusions}
\label{sec:conclusions}
In this study, we explored how magnetic activity affects mean density and acoustic radius inversions of the Sun as a star. Section~\ref{sec:datasets_and_modelling_strategy} outlines our modelling strategy, while Sect.~\ref{sec:influence_magnetic_activity_cycle} quantifies the impact of magnetic activity using Doppler velocity observations by a space-based instrument, GOLF, and a ground-based network, BiSON, and spanning two full solar cycles. In Sect.~\ref{sec:discussion}, we meticulously estimated various sources of systematic uncertainty influencing the inverted mean density and acoustic radius. We focused particularly on uncertainties related to the choice of physical ingredients in stellar models, stellar activity, and the surface effect prescription. Additionally, we evaluated the performance of the inversions conducted in this study and provided a table with recommended values and suggested uncertainties to be adopted.

Our research has demonstrated a non-negligible impact of the solar magnetic activity cycle on the asteroseismic characterisation of the Sun using inverse methods. This result complements the findings of \citet{Betrisey2024_MA_Sun}, who reached the same conclusion using forward methods. This impact was evident in two independent datasets, GOLF and BiSON, and persisted even when the method for estimating the surface effect coefficients and the surface effect prescription were modified. For the mean density, we observed an average uncertainty of 0.15\% and 0.19\% due to magnetic activity for GOLF and BiSON data, respectively. These values are substantial, as they exceed the statistical uncertainty and represent the largest source of systematic uncertainty. Similar results were found for the acoustic radius, with average uncertainties of 0.026\% and 0.033\% for GOLF and BiSON data, respectively.

Consistent with the findings of \citet{Betrisey2024_MA_Sun}, we find that including low radial-order modes in the datasets mitigates the effects of magnetic activity. Unlike the imprint on stellar age obtained by \citet{Betrisey2024_MA_Sun} through forward modelling, this mitigation is significant. Specifically, the imprint can be reduced by a factor of approximately 1.5 to 2 if low-order modes are included. This is promising for the PLATO mission, as the mean density and acoustic radius are part of the quantities of interest of the mission. However, detecting these oscillations will be challenging, as PLATO measures photometric variations \citep[e.g.][]{Rauer2024} rather than the radial velocity variations used in this study. Nonetheless, it is conceivable that some of the best PLATO observations might achieve the precision required for such measurements \citep{Goupil2024}, similar to the \textit{Kepler LEGACY} sample \citep{Lund2017}, which originates from a similar observing strategy.

In the case of the Sun, magnetic activity is the largest source of systematic uncertainty, followed by the choice of physical ingredients in stellar models and the choice of the surface effect prescription. These two latter factors are however calibrated to reproduce the solar properties, and further investigations are therefore needed to assess their impact on more massive stars with convective cores, such as F-type stars. Including all systematics, we obtained final recommended values of $\bar{\rho}_{\mathrm{inv}} = 1.4104\pm 0.0051$ g/cm$^3$ and $\tau_{\mathrm{inv}} = 3722.0\pm 4.1$ s.

Putting these results in the context of the PLATO mission, we achieved a high precision of 0.36\% and 0.11\% for mean density and acoustic radius, respectively. This high precision includes the major sources of systematic uncertainty. Additionally, although the robustness of mean density and acoustic radius inversions has been extensively demonstrated in the literature \citep[e.g.][]{Reese2012,Buldgen2015a}, our study further demonstrates this robustness. These results are encouraging, as they indicate the potential to reach high levels of precision for Sun-like stars on these quantities. A better-constrained mean density can be used as an additional constraint along with a fit of frequency separation ratios, significantly improving the precision of the stellar radius \citep[e.g.][]{Buldgen2019f,Betrisey2022,Betrisey2023_AMS_surf}. This improvement is crucial for characterising exoplanetary systems, as a better-constrained stellar radius enhances the determination of the orbital distance and planetary radius of exoplanets detected by the transit method. Based on the results of this study and of \citet{Betrisey2022}, who examined a host star with properties similar to the PLATO benchmark of Sun-like stars, we believe that it will not be rare to achieve this precision for future observations of Sun-like stars by PLATO. Further investigations are however needed for targets more active than the Sun, as the impact of magnetic activity on stellar characterisation is greater for these targets \citep{Betrisey2024_MA_Sun}. Extending these investigations to more massive stars with convective cores is also of prime importance. These stars are expected to constitute a significant fraction of future PLATO observations \citep[e.g.][]{Goupil2024}, therefore requiring to develop a map of the parameter space and provide prescriptions for systematic uncertainties that could be incorporated into the modelling pipeline, rather than requiring detailed modelling of each target to quantify these uncertainties.

\begin{acknowledgements}
J.B. acknowledges funding from the SNF Postdoc.Mobility grant no. P500PT{\_}222217 (Impact of magnetic activity on the characterization of FGKM main-sequence host-stars). S.N.B acknowledges support from PLATO ASI-INAF agreement no. 2022-28-HH.0 "PLATO Fase D". R.A.G. acknowledges the support from PLATO and GOLF CNES grants. A.M.A acknowledges support from the Swedish Research Council (VR 2020-03940) and from the Crafoord Foundation via the Royal Swedish Academy of Sciences (CR 2024-0015). A.-M.B. has received support from STFC consolidated grant ST/X000915/1. O.K. acknowledges support by the Swedish Research Council (grant agreements no. 2019-03548 and 2023-03667), the Swedish National Space Agency, and the Royal Swedish Academy of Sciences. Finally, this work has benefited from financial support by CNES (Centre National des Études Spatiales) in the framework of its contribution to the PLATO mission.
\end{acknowledgements}

\bibliography{bibliography.bib}

\appendix
\onecolumn

\section{Consistency tests}
\label{sec:preliminary_tests}
\begin{figure*}[h!]
\resizebox{0.99\linewidth}{!}{
\begin{subfigure}[b]{.49\textwidth}
  \includegraphics[width=.99\linewidth]{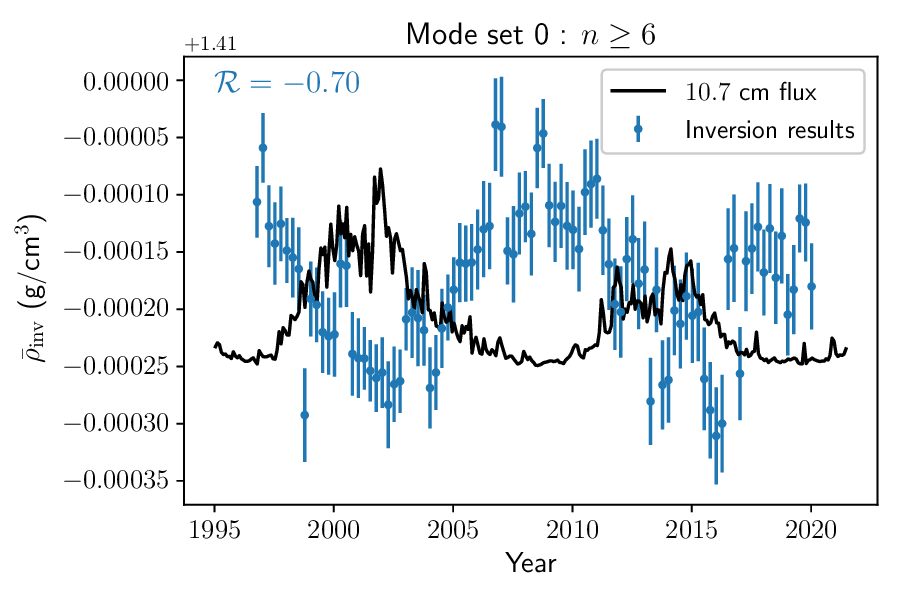}  
\end{subfigure}
\begin{subfigure}[b]{.49\textwidth}
  \includegraphics[width=.99\linewidth]{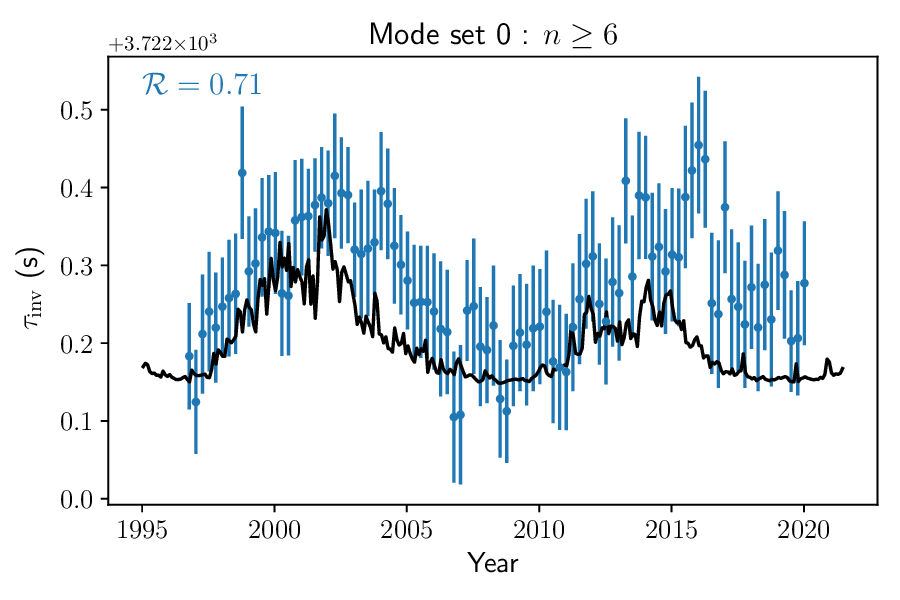} 
\end{subfigure}}
\resizebox{0.99\linewidth}{!}{
\begin{subfigure}[b]{.49\textwidth}
  \includegraphics[width=.99\linewidth]{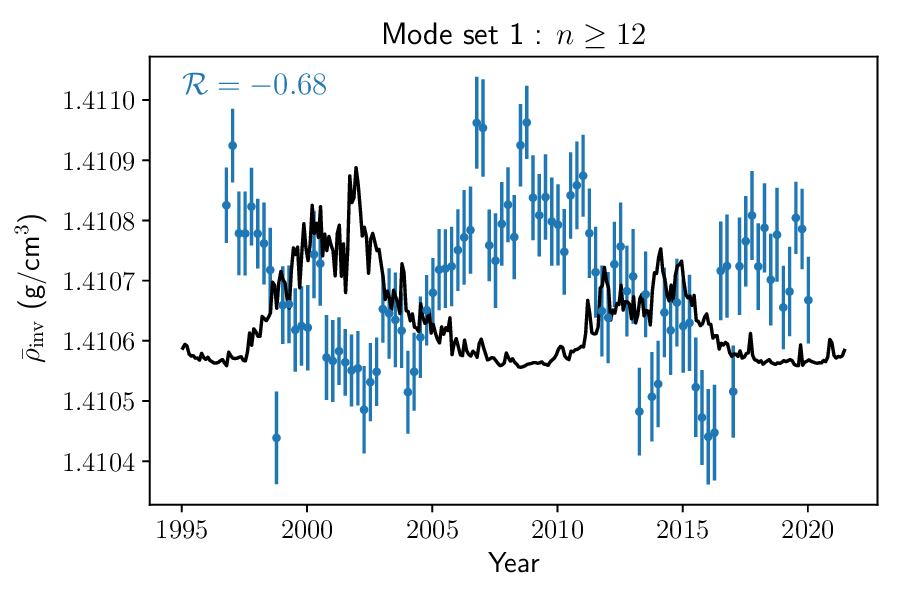}  
\end{subfigure}
\begin{subfigure}[b]{.49\textwidth}
  \includegraphics[width=.99\linewidth]{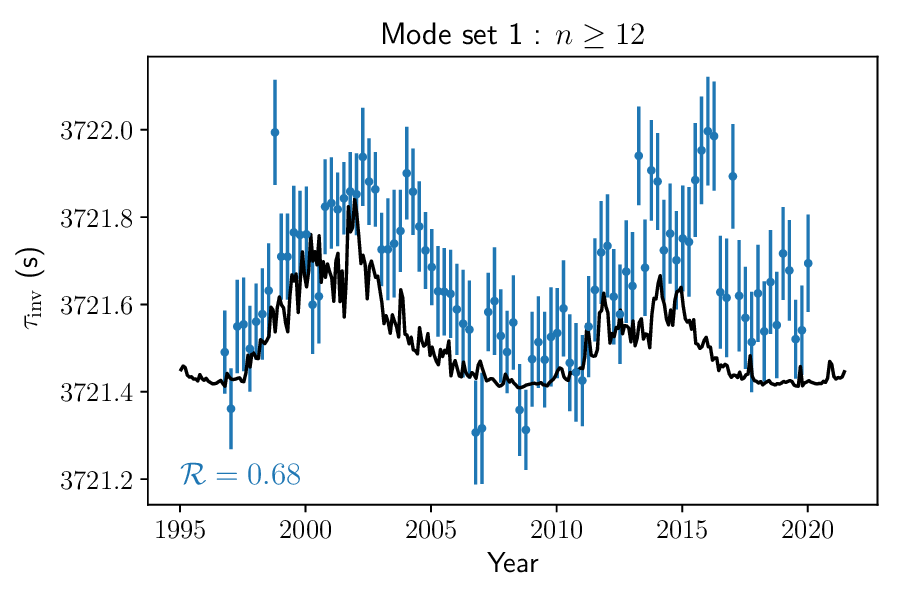} 
\end{subfigure}}
\resizebox{0.99\linewidth}{!}{
\begin{subfigure}[b]{.49\textwidth}
  \includegraphics[width=.99\linewidth]{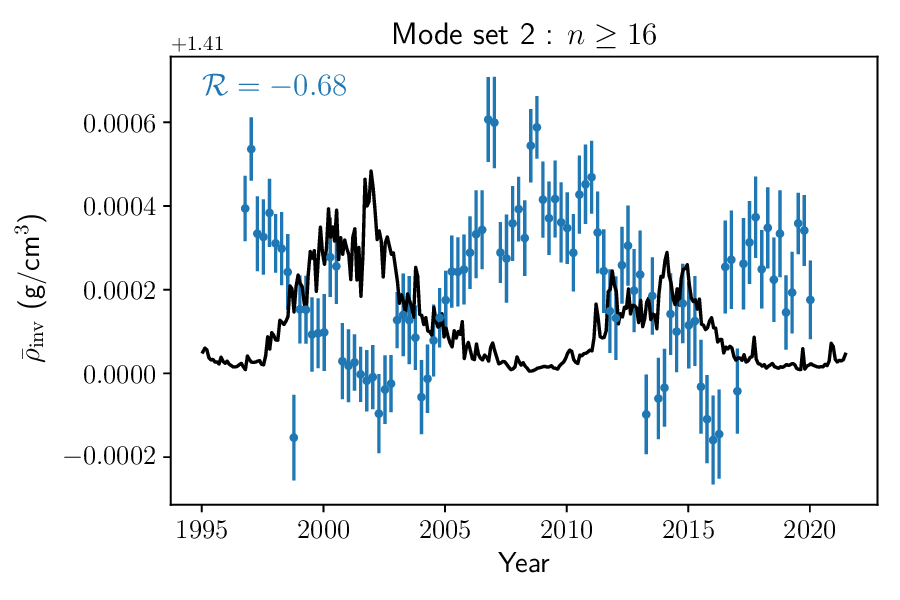}  
\end{subfigure}
\begin{subfigure}[b]{.49\textwidth}
  \includegraphics[width=.99\linewidth]{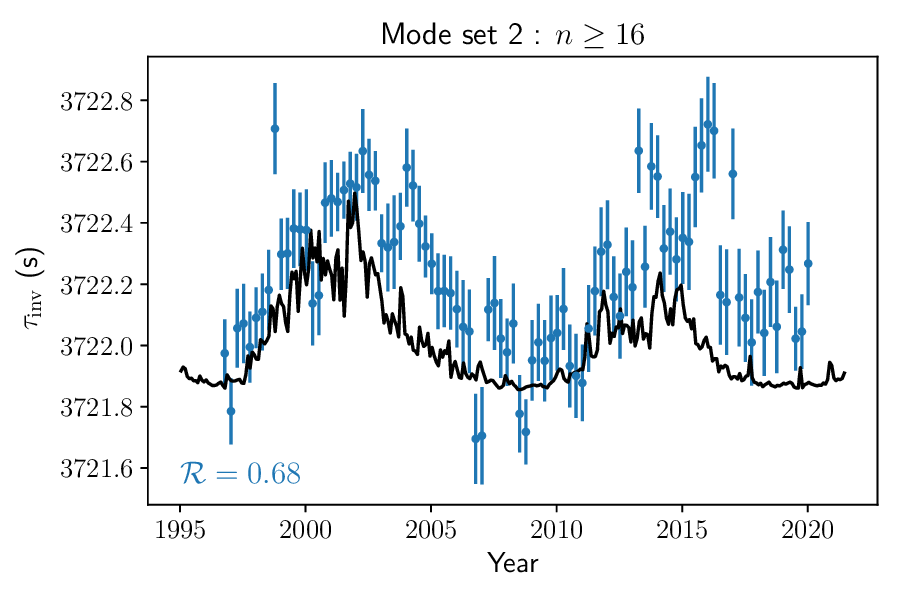}  
\end{subfigure}}
\caption{Results of the consistency tests assessing the effectiveness of the automation of the InversionKit software for the mean density (left column) and acoustic radius (right column) inversions. The number $\mathcal{R}$ denotes the Spearman correlation coefficient between the dataset (in blue) and the activity proxy (solid black line), the 10.7 cm radio emission flux. The activity proxy was rescaled for illustration purposes.}
\label{fig:results_preliminary_tests}
\end{figure*}

To evaluate the reliability of the automated version of InversionKit, we began by comparing its results with those from the previous version designed for individual modelling. We confirmed that both versions produced identical outcomes. To further mitigate potential systematic errors, we conducted additional tests. First, we performed an inversion on the same reference model, artificially introducing cycle-dependent surface effect coefficients and verifying that this signal is indeed propagated to the inversion results. To that end, we used the cycle-dependent coefficients obtained by \citet{Betrisey2024_MA_Sun} from GOLF data for the set labelled `BG2,~$n \geq 12$' in their article. We then repeated this process for mode sets~0 and~1, which include additional low radial-order oscillation modes. Given the forced introduction of the cycle-dependent signal, we anticipated that modes less sensitive to surface effects and magnetic activity would produce a reduced impact on the results. The impact on the inversion results was assessed using two measures: the standard deviation of the inversion results and the difference between the maximum and minimum values, which respectively underestimate and overestimate the quantitative impact of the cycle (see Sect.~\ref{sec:systematic_uncertainties}). The results, presented in Fig.~\ref{fig:results_preliminary_tests} and Table~\ref{tab:results_percentages_preliminary}, confirmed that the imprint of the activity cycle was consistently propagated to the inversion results, independently of the configuration that was tested. As expected, the impact was smallest for set~0, larger for set~1, and largest for set~2.

\begin{table}[h!]
\centering
\caption{Results of the consistency tests assessing the effectiveness of the automation of the InversionKit software.}
\begin{tabular}{lcccc}
\hline \hline
\multirow{2}{*}{Mode set} & \multicolumn{2}{c}{$\bar{\rho}_{\mathrm{inv}}$} & \multicolumn{2}{c}{$\tau_{\mathrm{inv}}$} \\ \cmidrule[0.4pt](lr){2-3} \cmidrule[0.4pt](l){4-5}
 & std & min/max & std & min/max \\ 
\hline 
Set 0: $n \geq 6$ & 0.004\% & 0.019\% & 0.002\% & 0.009\% \\ 
Set 1: $n \geq 12$ & 0.009\% & 0.037\% & 0.004\% & 0.019\% \\ 
Set 2: $n \geq 16$ & 0.013\% & 0.054\% & 0.006\% & 0.028\% \\ 
\hline 
\end{tabular}
{\par\small\justify\textbf{Notes.} The `std' column represents the standard deviation of the dataset, expressed as a percentage of the mean value. The `min/max' column shows the difference between the maximum and minimum values, also expressed as a percentage of the mean value of the dataset. \par}
\label{tab:results_percentages_preliminary}
\end{table}

\section{Performance evolution of GOLF and BiSON}
\label{app:supplementary_data}
\begin{figure}[h!]
\begin{subfigure}[b]{.49\textwidth}
  \includegraphics[width=.99\linewidth]{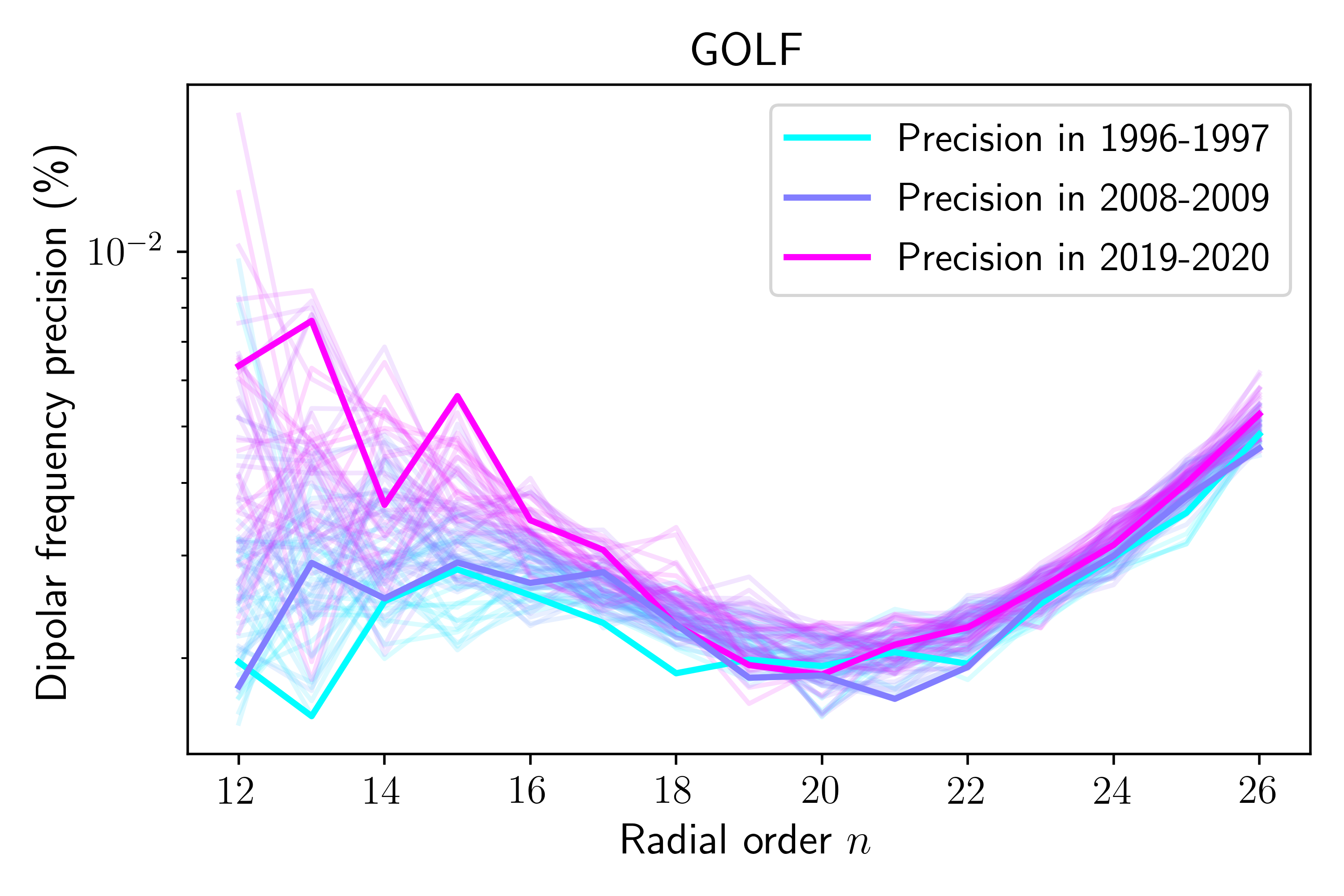}  
\end{subfigure}
\begin{subfigure}[b]{.49\textwidth}
  \includegraphics[width=.99\linewidth]{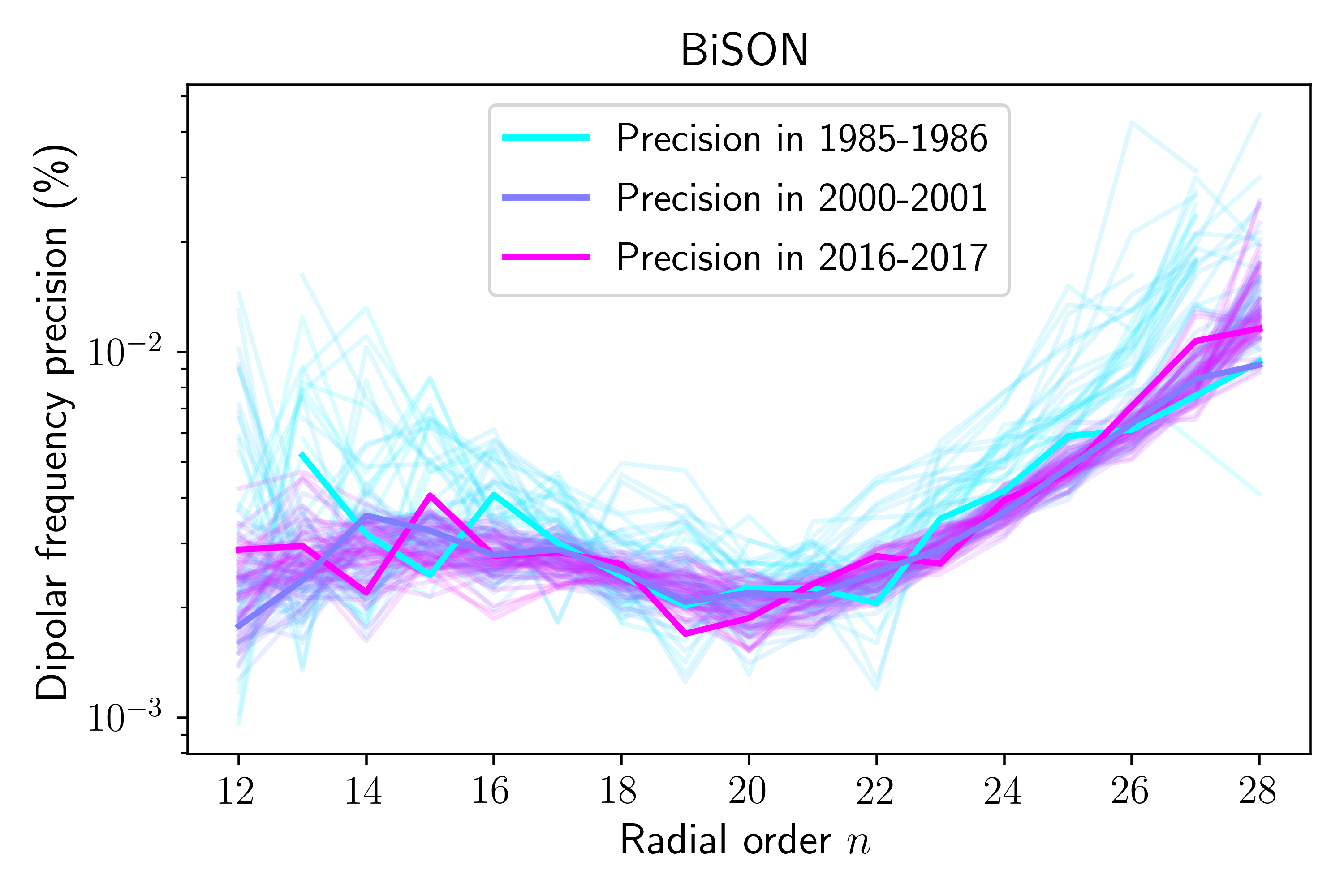} 
\end{subfigure}
\caption{Evolution of the observational precision achieved for frequency of dipolar modes detected by the GOLF instrument (upper panel) and the BiSON network (lower panel). Each line corresponds to a yearly snapshot and the color map transitions from bluish colors for the early instrument years to pinkish colors for the recent years.}
\label{fig:appendix_evol_precision_comp}
\end{figure}

In Fig.~\ref{fig:appendix_evol_precision_comp}, we depict the evolution of observational precision for frequency of dipolar modes detected by GOLF and BiSON. Each line represents a yearly snapshot, staggered by three months. The GOLF instrument \citep{Gabriel1995}, which has been orbiting the Sun aboard the SoHO spacecraft \citep{Domingo1995} since its launch in the mid-1990s, could not been maintained, leading to a decline in performance as the instrument ages. Specifically, the background photon noise contribution increases in the power spectrum, making it increasingly difficult to detect frequencies distant from the frequency of maximal power \citep{Garcia2005}. In our case, this degradation particularly affects frequencies in the radial order range $n=12-15$, as indicated by the pink line in the upper panel of Fig.~\ref{fig:appendix_evol_precision_comp}.

In contrast, the BiSON network \citep{Davies2014,Hale2016} consists of ground-based facilities that can be maintained over time. The lowest data quality was observed in the mid-1980s, as shown by the blue lines in the lower panel of Fig.~\ref{fig:appendix_evol_precision_comp}, when only one observing node was operational. The data quality has since improved with the addition of more observing nodes to the BiSON network.

\end{document}